\documentstyle[sprocl]{article}

\def\be{\begin{equation}}
\def\ee{\end{equation}}
\def\bea{\begin{eqnarray}}
\def\eea{\end{eqnarray}}

\begin{document}

\title{HOW TO EXPLICITLY INTRODUCE THE JAFFE-WITTEN MASS GAP INTO THE QUANTUM  
YANG-MILLS THEORY}

\author{V. GOGOHIA}

\address{HAS, CRIP, RMKI, Theory Division, 
P.O.B. 49, H-1525 Budapest 114, Hungary \\ gogohia@rmki.kfki.hu } 




\maketitle\abstracts{We propose how to explicitly introduce the Jaffe-Witten   
mass gap into the quantum Yang-Mills theory. Through the full gluon propagator 
it is defined as such nonperturbative scale that when it formally    
goes to zero, then the perturbative phase survives in the theory only.        
The close link between mass gap and strong infrared sigularities which are 
due to dominated in the QCD vacuum self-interaction of massless gluons is also 
discussed. This interaction leads thus to the zero momentum modes enhancement  
effect in the QCD NP vacuum. Using theory of distributions, we argue that
strong infrared singularities can be put under control. A new, intrinsically   
nonperturbative phase in QCD is established. }

\section{Introdiction}

The system of quantum, dynamical equations of motion, the so-called
Schwinger-Dyson (SD) equations for lower (propagators) and higher
(vertices and kernels) Green's functions [1] can serve as an adequate and      
effective tool for the nonperturbative (NP) approach to QCD. However, to say   
today that QCD is the NP theory is almost tautology. The problem is  
how to define it exactly since we surely know that QCD has a perturbative (PT)
phase as well because of asymptotic freedom (AF) [1]. 
In order to exactly define NP QCD, let us first start from the Yang-Mills (YM) 
sector. The two-point Green's function, describing the full gluon    
propagator, is

\begin{equation}
iD_{\mu\nu}(q) = \left\{ T_{\mu\nu}(q)d(-q^2, \xi) + \xi L_{\mu\nu}(q) \right\} {1 \over q^2 }.
\end{equation}
Here $\xi$  is a gauge fixing parameter and         
$T_{\mu\nu}(q)=g_{\mu\nu}-(q_{\mu} q_{\nu} / q^2) = g_{\mu\nu } - L_{\mu\nu}(q)$. For its free (tree level) counterpart see Eq. (8) below. The solutions of   
the corresponding SD equation for the full gluon propagator (1)
are supposed to reflect the complexity of the quantum structure
of the QCD true ground state. It is a highly nonlinear system of      
four-dimensional integrals containing many different, unknown in general       
propagators, vertices and kernels, so there is no hope for exact solution(s).  
In any case, however, the solutions of this equation can be distinguished by
their behavior in the deep infrared (DIR) limit, describing thus many
(several) different types of quantum excitations and fluctuations of gluon
field configurations in the QCD NP vacuum.                                     
Evidently, not all of them can reflect its real structure.
 
The realistic ultraviolet (UV)    
limit of these solutions is uniquely determined since we know solutions of the 
renormgroup equations because of AF. At the same time, the solutions of the renormgroup equations in the DIR region is not known. That is why in 
this case one has to rely on the formal asymptotics of the full gluon          
propagator as they follow from its expression (1).    
The DIR asymptotics of the full gluon propagator can be generally classified 
into the two different types: singular which means the IR enhanced (IRE) or   
smooth which means the IR finite (IRF) or even the IR vanishing (IRV) gluon    
propagators. However, the smooth behavior of the full gluon propagator is      
possible only in one exceptional gauge - the Landau gauge ($\xi = 0$), i.e., it
is a gauge artefact. We will not therefore discuss it in what follows, so we   
are left with the IRE one only.       
                                                   
Let us now emphasize that any deviation in the behavior of the full
gluon propagator in the DIR domain from the free one
automatically assumes its dependence on a scale parameter (at
least one) in general different from QCD asymptotic scale
parameter $\Lambda_{QCD}$. It can be considered as responsible
for the NP dynamics (in the IR region) in the QCD
vacuum. If QCD itself is a confining theory, then   
such a characteristic scale is very likely to exist. In what follows,
let us denote it, say, $\Delta$ (see below).

\section{Jaffe-Witten mass gap}

{\bf Truly NP phase.} The phenomenon of ''dimensional    
transmutation'' [1] only supports our general conclusion that QCD exhibits  
a mass determining the scale of NP dynamics in its ground state. In terms    
of the full gluon propagator it can be exactly defined as follows:             
                                 
\begin{equation}
d^{TNP}(-q^2, \Delta^2) = d(-q^2, \Delta^2) - d(-q^2, \Delta^2 =0),
\end{equation}
where we introduce explicit dependence on the above-mentioned NP mass scale    
parameter $\Delta$. This subtraction can be also considered as the    
definition of the truly NP (TNP) part
of the full gluon propagator since when the NP scale parameter goes formally   
to zero then the TNP part vanishes.        
The definition (2) also explains the difference between the 
TNP part $d^{TNP}(-q^2)$ and the full gluon propagator $d(-q^2)$ which is NP   
itself. On the other hand, the subtraction (2) is equivalent to                
                           
\begin{equation}
d^{TNP} (- q^2, \Delta^2) = \Delta^2 f(- q^2, \Delta^2),
\end{equation}
where the function $f(q^2, \Delta^2)$ is of the corresponding dimension and it 
has a finite limit as $\Delta^2 \rightarrow 0$. Thus in the initially massless 
quantum YM theory the characteristic mass appears explicitly. In Ref. [2] Jaffe
and Witten (JW) have argued that quantum YM fields should exhibit              
characteristic mass scale   
the so-called mass gap $\Delta$ in order to explain why the nuclear force is   
strong but short-range. The definitions (2) and (3) is precisely our proposal  
how the JW mass gap could be explicitly introduced into the quantum YM theory.
In other words, by defining TNP QCD, we identify the NP scale with their mass  
gap. Let us note further that the
limit $\Delta^2 \rightarrow 0$ is usually equivalent to the UV
limit $-q^2 \rightarrow \infty$, so it is almost obvious that because of AF    
one can identify

\begin{equation}
d(-q^2, \Delta^2 =0) \equiv d^{PT}(-q^2).
\end{equation} 
Thus the relation (4) 
is our definition of the PT phase in QCD, then the relation (2) becomes
 
\begin{equation}
d(-q^2, \Delta^2) = d^{TNP}(-q^2, \Delta^2) + d^{PT}(-q^2).
\end{equation}
Let us underline that this is an exact relation (the gluon momentum runs
over the whole range $[0, \infty)$), so at this stage there is no      
approximation made. Due to the inevitable identification (4), one can say 
that we define the NP scale parameter in the way that when it formally goes    
to zero (i.e., when the TNP part vanishes, Eq. (3)), then the PT phase only    
survives in the full gluon propagator.
This allows one to clearly separate the JW mass gap from        
all other mass scale parameters which may be present in the
full gluon propagator. It is worth emphasizing that in the realistic   
models  of the full gluon propagator its TNP part usually
coincides with its DIR asymptotics determining thus the strong
intrinsic influence of the IR properties of the theory on its
NP dynamics.

Substituting the exact decomposition (5) into the full gluon propagator
(1), one obtains

\begin{equation}
D_{\mu\nu}(q, \Delta) = D^{TNP}_{\mu\nu}(q, \Delta) + D^{PT}_{\mu\nu}(q),
\end{equation}
where                                       

\begin{equation}
 D^{TNP}_{\mu\nu}(q,\Delta) = - i T_{\mu\nu}(q) d^{TNP}(-q^2, \Delta^2){ 1 \over q^2}, 
\end{equation}

\begin{equation}
D^0_{\mu\nu}(q) = - i \{ T_{\mu\nu}(q) + \xi L_{\mu\nu}(q) \} { 1 \over q^2},
\end{equation}

\begin{equation}
D^{PT}_{\mu\nu}(q) = - i \{ T_{\mu\nu}(q)  d^{PT}(-q^2) + \xi L_{\mu\nu}(q) \}
{ 1 \over q^2}.
\end{equation}
For further purposes the explicit expression for the free gluon          
propagator $D^0_{\mu\nu}(q)$ is also given.  
Thus the exact decomposition (6) has a few remarkable features. First of all,  
the dependence of    
the full gluon propagator on the JW mass gap $\Delta$ is exactly placed in its 
TNP part. Secondly, the explicit gauge dependence of the full      
gluon propagator is exactly shifted from its TNP part to its PT part (for      
reasons to proceed in this way, see discussion below). Thus we clearly separate
the NP phase from the PT one in QCD.

{\bf Intrinsically NP phase.} Up to this moment we have delt only with exact decompositions and definitions, i.e., only algebraic manipulations have been     
done. It is the time now to introduce nontrivial dynamics into this scheme.    
Evidently, the only place  
where it can be done is, of course, the DIR region, i.e., by saying something  
nontrivial about the DIR asymptotics of the full gluon propagator one can      
additionally  distinguish between its TNP and PT parts. Fortunately, we have  
an exact criterion for establishing the DIR structure of the full gluon        
propagator. The free gluon propagator (8)    
has an exact $(1/q^2)$ singularity in the DIR limit since it is          
defined in the whole range. The PT part, presented by Eq. (9) 
also in the whole range, possesses the same property.  This means that the  
free gluon singularity $(1/q^2)$ at $q^2 \rightarrow 0$ is an exact separation 
line between the TNP and PT parts in the full gluon propagator (6). It is worth
reemphasizing that decomposition (6) is exact, i.e., the gluon momentum runs over the whole range in both terms in its right hand side. The PT phase in QCD   
can be defined as one which in the IR is singular as the free gluon propagator.
Its existence in QCD is important from conceptual point of view and it is      
determined by AF. 

Contrary to the PT part, the TNP part of the full gluon propagator (7)
should have then singularities in the DIR domain stronger than $(1/q^2)$. Of   
course,
such strong singularities can be only of dynamical origin. The only dynamical 
mechanism in QCD which can produce such severe singularities in the vacuum is  
self-interaction of massless gluons in the DIR domain.                    
Precisely this self-interaction in the UV limit leads to AF. Thus one comes to 
the inevitable conclusion that the TNP part of the full gluon propagator should
have strong IR singularities different from those of the PT part.              
In what follows this type of singularities will be called as
NP IR singularities. Thus they should be summarized by the full gluon propagator and effectively correctly described by its TNP part in the DIR domain. In
this way the TNP phase becomes intrinsically NP (INP) one. So the definition of
the INP YM quantum theory consisits of the two conditions.  

I). The first necessary condition defines the TNP part of the full gluon       
propagator (7), on account of the definitions (2) and (3).

II). The second sufficient condition specifies the existence and structure of  
more stronger than $(1/q^2)$ IR singularities in the QCD NP vacuum.

The longitudinal part of the full gluon propagator is exactly singular in the  
IR as $(1/q^2)$, and therefore it does not make  
any sense to decompose the gauge fixing paramete $\xi$ similar to Eq. (2),     
though formally it is possible to do, of course.    
Thus the INP part of the full gluon propagator is manifestly gauge invariant   
and only transfer (physical) degrees of freedom of the gauge bosons are        
important for the NP dynamics in QCD.

\section{ZMME quantum model of the QCD ground state}

The quantum structure of the QCD NP vacuum is dominated by such types of       
excitations and fluctuations of gluon field configurations there which are due 
to self-interaction of massless gluons since precisely this interaction is 
the main quantum, dynamical effect in QCD.
In the UV region it implies AF. In the DIR region it becomes strongly singular 
and thus to be responsible for the zero momentum modes enhancement (ZMME)    
effect in the QCD NP vacuum. In the quantum YM theory all the NP IR singularities should be absorbed into the full gluon propagator, in its INP part. 
Thus the ZMME (or simply zero modes           
enhancement (ZME) since we work always in the momentum space) quantum,         
dynamical model of the QCD true ground state is based on the 
existence and importance of such kind of the NP excitations and fluctuations   
of gluon field configurations there which are due to self-interaction of massless gluons only (without explicitly involving some extra degrees of freedom).   
They are to be summarized by the INP part of the full gluon propagator and are 
to be effectively correctly described by its behavior in the DIR domain. 

In general, all the Green's functions in QCD are generalized functions, i.e.,  
they are distributions. Especially this is true for the severe NP IR      
singularities due to self-interaction of massless gluons in the QCD vacuum.    
Not loosing generality, the severe NP IR         
singularities can be analytically taken into account  
in terms of the TNP gluon form factor in Eq. (7) with a Euclidean signature  
($-q^2 \rightarrow q^2$) as follows:

\begin{equation}
d^{INP}(q^2, \Delta^2)= (\Delta^2)^{-\lambda - 1} (q^2)^{\lambda} \times f(q^2),
\end{equation}
where obviously we include $1/q^2$ from Eq. (7) into the exponent $\lambda$
which in general is arbitrary (any complex number with $Re \lambda < 0$, see   
below).
The function $f(q^2)$ is a dimensionless one  
which is regular at zero and otherwise remaining arbitrary, but preserving 
AF in the UV limit. This is nothing else but the analytical         
formulation of the second sufficient condition of the existence of the INP     
phase in the YM theory. That is why we replaced
superscipt ``TNP'' by ``INP'' in Eq. (10).
Since we are particulary interested in the DIR region, the arbitrary function 
$f(q^2)$ should be also expanded   
around zero in the form of the Taylor series in powers of $q^2$.           
As a result, we will be left with finite sum  
of power terms with exponent decreasing by one starting from $- \lambda$. All
other remaining terms from the Taylor expansion starting from the term   
having already the PT IR singularity should be   
shifted to the PT part of the full gluon propagator.                           
                                     
The distribution theory (DT) [3] tells us that         
the distribution $(q^2)^{\lambda}$ will have a simple pole at points  
$\lambda = - (n/2) - k, \ (k=0, 1, 2, 3...)$, where $n$ denotes the number    
of dimensions in Euclidean space ($q^2 = q^2_0 + q^2_1 + q^2_2 + ... + q^2_{n-1}$). In order to actually define the system of SD equations in the DIR
domain, it is convenient to introduce the IR regularization parameter   
$\epsilon$, defined as $D = n + 2 \epsilon, \ \epsilon \rightarrow 0^+$       
within a gauge invariant dimensional regularization (DR) method [1]. As a      
result, all the Green's functions should be regularized 
with respect to $\epsilon$ which is to be set to zero at   
the end of computations. The structure of the NP IR singularities is then      
determined (when $n$ is even number) as follows [3,4]:   
                                                    
\begin{equation}
(q^2)^{\lambda} = { C_{-1}^{(k)} \over \lambda +(D/2) + k} + finite \ terms,
\end{equation}
where the residue is 

\begin{equation}
 C_{-1}^{(k)} = { \pi^{n/2} \over 2^{2k} k! \Gamma ((n/2) + k) } \times        
L^k \delta^n (q)
\end{equation}
with $L = (\partial^2 / \partial q^2_0) + (\partial^2 / \partial q^2_1)
+ ... + (\partial^2 / \partial q^2_{n-1})$. In terms of $\epsilon$ from        
Eq. (11), one has                                          

\begin{equation}
(q^2)^{- {D \over 2} - k + \epsilon} = { 1 \over \epsilon} C_{-1}^{(k)} +      
finite \ terms,
\end{equation}
where we can put $D=n$ now. This means that the 
order of singularity does not depend on $\lambda$, i.e., it is always a simple 
pole, $(1/ \epsilon)$. Thus the wide-spread opinion that severe IR             
singularities (stronger than $(1/q^2)$) in QCD NP vacuum cannot be controlled  
is not justified (for details see Refs. [3,4]).

\end{document}